\documentclass[preprint,12pt]{elsarticle}

\usepackage{amssymb}

\usepackage{lineno}

\begin{document}

\begin{frontmatter}



\title{Results and prospects of dark matter searches with ANTARES}


\author[IFIC]{J. D. Zornoza\corref{Corresponding author}\fnref{label2}}
\author[IFIC]{G. Lambard}
\fntext[label2]{Corresponding author: zornoza@ific.uv.es}

\address[IFIC]{IFIC, Instituto de F\'{i}sica Corpuscular (CSIC-Universidad de
Valencia, Ed. Institutos de Investigaci\'{o}n, AC22085, E46071, Valencia, Spain}

\author{on behalf of the ANTARES Collaboration}

\begin{abstract}
  Dark matter is one of the most important scientific goals for
  neutrino telescopes. These instruments have particular advantages
  with respect to other experimental approaches. Compared to direct
  searches, the sensitivity of neutrino telescopes to probe the
  spin-dependent cross section of WIMP-proton is unsurpassed. On the
  other hand, neutrino telescopes can look for dark matter in the Sun,
  so a potential signal would be a strong indication of dark matter,
  contrary to the case of other indirect searches like gammas or
  cosmic rays, where more conventional astrophysical interpretations
  are very hard to rule out. We present here the results of a binned
  search for neutralino annihilation in the Sun using data gathered
  by the ANTARES neutrino telescope during 2007-2008. These result
  include limits on the neutrino and muon flux and on the
  spin-dependent and spin-independent cross section of the WIMP-proton scattering.

\end{abstract}

\begin{keyword}
dark matter \sep WIMP \sep neutralino \sep neutrino telescopes


\end{keyword}

\end{frontmatter}


\section{Introduction}
\label{intro}

Dark matter existence has been soundly proofed by different
experimental evidence, including the observations from
Planck~\cite{planck}, the results on the Big Bang
Nucleosynthesis~\cite{jedamzik}, the rotation curves of
galaxies~\cite{rubin} and the studies of highly red-shifted Ia
supernovae~\cite{kowalski}. These results show that the only about
30\% of the content of the Universe is matter and about 70\% is dark
energy. Moreover, 85\% of the matter is non-barionic. Explanations for
the nature of this non-barionic component have to be outside the
Standard Model. The basic conditions required to a particle dark
matter candidate are to have interaction cross section of the order of
that of the weak interaction and to be massive and stable. Particles
like neutrinos, which fulfill these requirement for a good dark matter
candidate, are not viable as a dominant component, since they are
relativistic and cannot explain the large-scale structure of the
Universe. Given these contiditons, a generic familly of particles
fulfilling these conditions are callend WIMPs (Weakly Interacting
Massive Particles). The most popular model which provides WIMP
candidates is Supersymmetry (SUSY). In particular, in this analysis we
have looked for neutralinos, which in many of the possible scenarios
is the lightest SUSY particle. Its stability is preserved by the
conservation of the R-parity. The results have been analyzed with
respect to two implementanions of the SUSY framework: CMSSM~\cite{cmssm} and MSSM-7~\cite{mssm7}.

The analysis presented here is a binned search for neutrinos produced
after the neutralino annihilations in the Sun direction using 2007-2008
data of the ANTARES neutrino telescope, since neutralinos would
accumulate in massive objects like the Sun and their
annihilation would produce high energy neutrinos~\cite{paper}. One of the
advantages of this kind of searches, compared to other indirect
searches like looking for gammas in the Galactic Center or excesses of
positrons is that a potential signal would be a very robust indication
of dark matter, since no other astrophysical explanations are expected.

The structure of this paper is as follows. The ANTARES detector is
introduced in Section~\ref{antares}. The estimations for
signal and background are described in
Section~\ref{simulation}. Section~\ref{cuts} explains the optimization
procedure. Finally, the results are presented in Section~\ref{results} and
the conclusions are summarized in Section~\ref{conclusions}.

\section{The ANTARES detector}
\label{antares}

The ANTARES neutrino telescope~\cite{antares} is located in the Mediterranean Sea, at
a depth of 2.5~km, about 40~km off Toulon, in the French coast. It
consists of 885 photomultipliers (PMTs) arranged in a three-dimensional
array. The operation principle is based on the detection by these PMTs of the
Cherenkov light induced by relativistic muons produced in CC
interaction of high energy neutrinos in the sourrondings of the
detector. The PMTs are installed along 12 lines anchored to the sea
floor and kept taut by buoys at the top of them. The length of the
lines is 450~m and the distance between lines is 60-75~m. The PMTs are
grouped in triplets in order to reduce the effect of optical
background produced by potassium-40 and bioluminescence. The position
and time information of the photons detected by the PMTs can be used
to reconstruct the muon direction.

The installation of the detector was completed in 2008, although during 2007 five lines
were already installed and produced data which can be used for physics analysis.

\section{Signal and background estimation}
\label{simulation}

The signal, i.e. the neutrino flux arriving at Earth from neutralino annihilations in
the Sun, is calculated using the WimpSim package~\cite{wimpsim}
for the relevant channels in this analysis ($q \bar{q}$, $l \bar{l}$,
$W^{+}$, $W^{-}$, ZZ, Higgs doublets $\phi \phi^{*}$ and $\nu
\bar{\nu}$)) for energies ranging from 10~GeV to 10~TeV. As benchmarks
for describing the range of potential signal, three channels have been
analized: $\rm \tilde{\chi}_{1}^{0}\tilde{\chi}_{1}^{0}\rightarrow
b\bar{b}$ (soft channel) and  $\rm
\tilde{\chi}_{1}^{0}\tilde{\chi}_{1}^{0} \rightarrow W^{+}W^{-}$ and $\rm \tilde{\chi}_{1}^{0}\tilde{\chi}_{1}^{0} \rightarrow \tau^{+}\tau^{-}$ (hard channels). In each
case, a 100\% branching ratio is assumed to represent the most extreme
cases in the parameter space. Oscillations among the three
neutrino flavours, $\nu$ absorption and $\tau$ regeneration in the
Sun's medium are taken into account by WimpSim.

There are two kinds of background for detectors like ANTARES. First
the muons produced by cosmic rays interacting in the atmosphere. In
order to reduce this kind of background, only upgoing event are
selected, since muons cannot traverse the Earth. However, some of
these muons (a small fraction of the total but a large number given
the flux involved) are misreconstructed as upgoing. This is why
further cuts are needed on the quality of the reconstructed track in
order to reject them. The second kind of background are the
atmospheric neutrinos, also produced in the cosmic ray interactions in
the atmosphere. This is an irreducible background. The background
estimation is done by scrambling real data, which reduces the impact
of systematic uncertainties.

\begin{figure}[c]
\begin{center}
\includegraphics[width=1.0\linewidth]{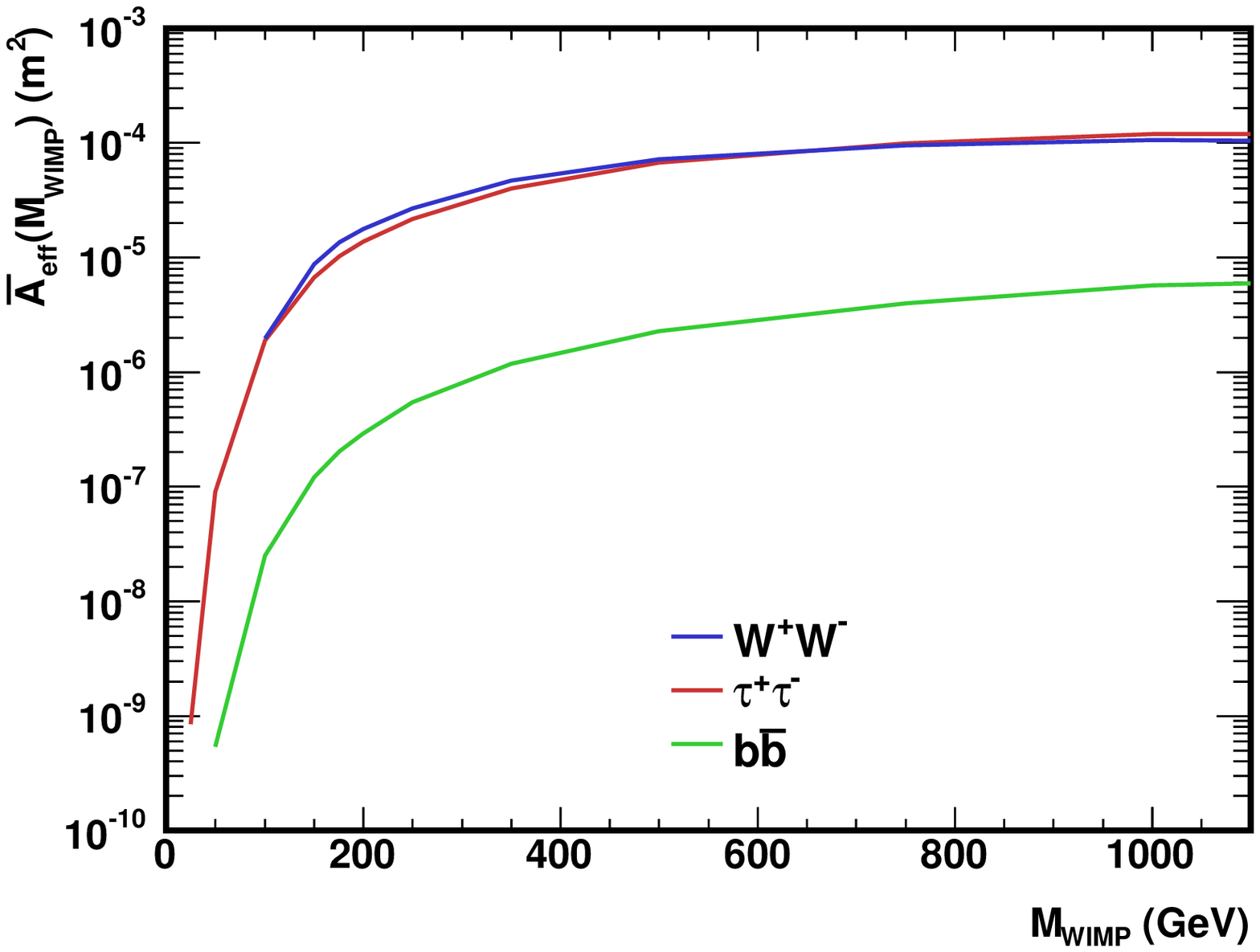}
\caption{Examples of the averaged effective area $\rm \bar{A}_{eff}(M_{\rm WIMP})$ 
for the signal of WIMP self-annihilation inside the Sun, $b\bar{b}$ (green), $W^{+}W^{-}$ (blue) and $\tau^{+}\tau^{-}$ (red) 
channels. The detector is in a 12 line configuration and the applied
cuts in the quality reconstruction parameter and the angular distance
with respect to the Sun are $\rm Q_{cut}<1.4$ and $\rm \Psi_{cut}<
3^{\circ}$, respectively. (Preliminary).}
\label{effarea}
\end{center}
\end{figure}

\section{Optimization}
\label{cuts}

The procedure for the analysis presented here has followed a blind
strategy in order to avoid selection biases. This means that the cuts
have been selected before looking at the source region. The criterion
to choose the cuts has been to optimize (minimize) the average upper
limit which can be set for any given neutralino mass. This average
upper limit can be calculated as

\begin{equation}
\rm{\overline{\Phi}_{\nu_{\mu}+\bar{\nu}_\mu} = \frac{\bar{\mu}^{90\%}}{\sum\limits_{i} \bar{A}_{eff}^{i}(M_{WIMP}) \times T_{eff}^{i}}} \, ,
\label{mrfeq}
\end{equation}

\noindent where $i$ refers to the different detector configurations
(5, 9, 10 and 12 lines), $\bar{\mu}^{90\%}$ is the average upper limit
at 90\% CL (calculated using Feldman-Cousins recipe~\cite{fc}) and
$T_{eff}^{i}$ is the livetime for each detector configuration. The
average effective area is defined as the equivalent area which would
be 100\% efficient to detect a neutrino flux and produce the same
number of events as the actual detector. Figure~\ref{effarea} shows
an example of the effective area for the studied channels.

The optimization is done by scanning the average upper limit for
different values of cuts in the search cone angle around the Sun
direction ($\Psi$) and the quality parameter assigned by the reconstruction
algorithm ($Q$) (~\cite{bbfit}). For each neutrino mass and channel, an
optimized set of cuts is found.

\section{Results}
\label{results}

\begin{figure}[c]
\begin{center}
\includegraphics[width=1.0\linewidth]{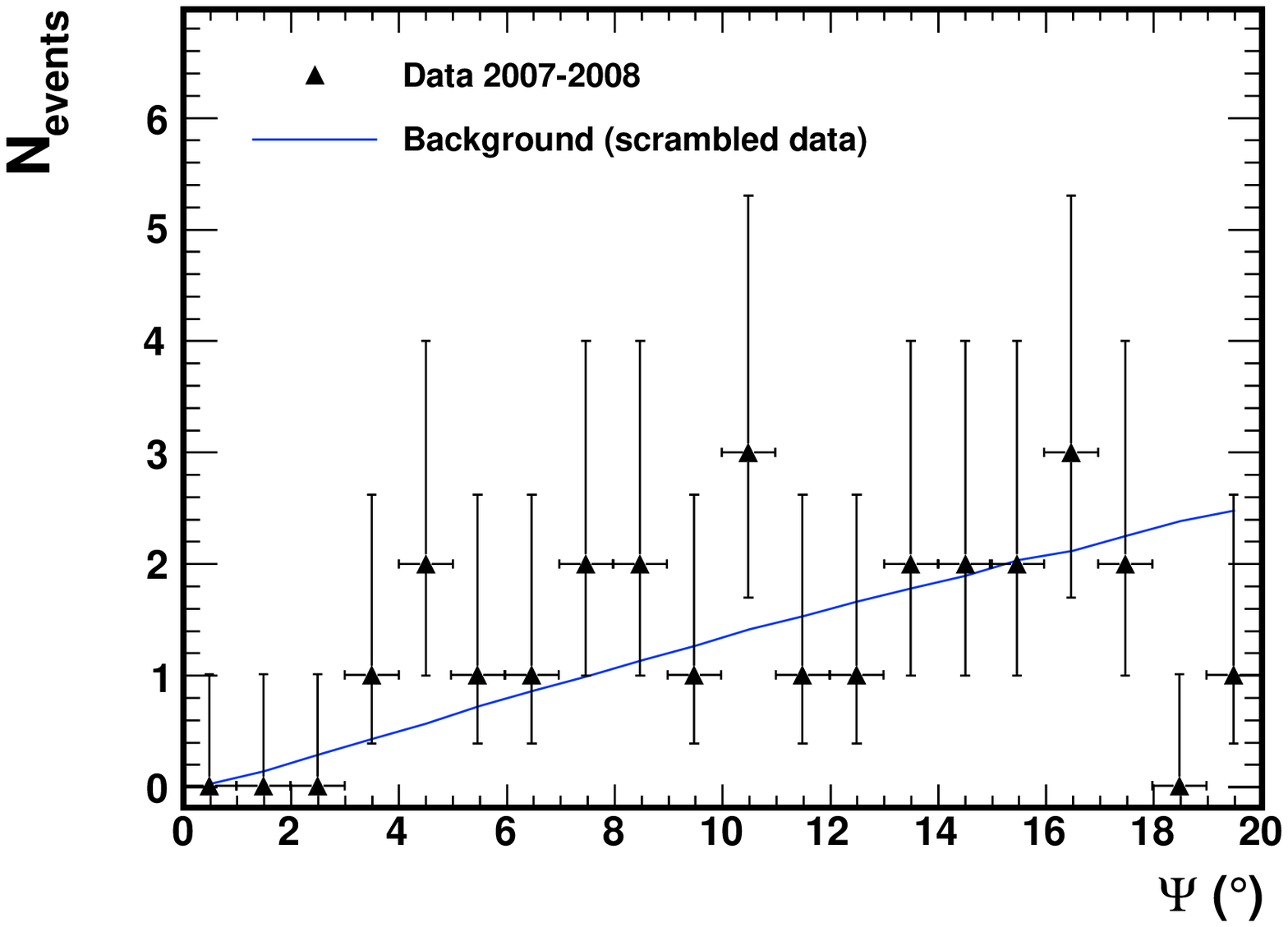}
\caption{Differential distribution of the angular separation $\rm \Psi$ of the event tracks 
with respect to the Sun's direction for the expected background (solid blue line) compared to 
the data (black triangles). A $1\sigma$ Poisson uncertainty is shown
for each data point. (Preliminary).}
\label{data}
\end{center}
\end{figure}

\begin{figure*}[c]
\begin{center}
\includegraphics[width=0.48\linewidth]{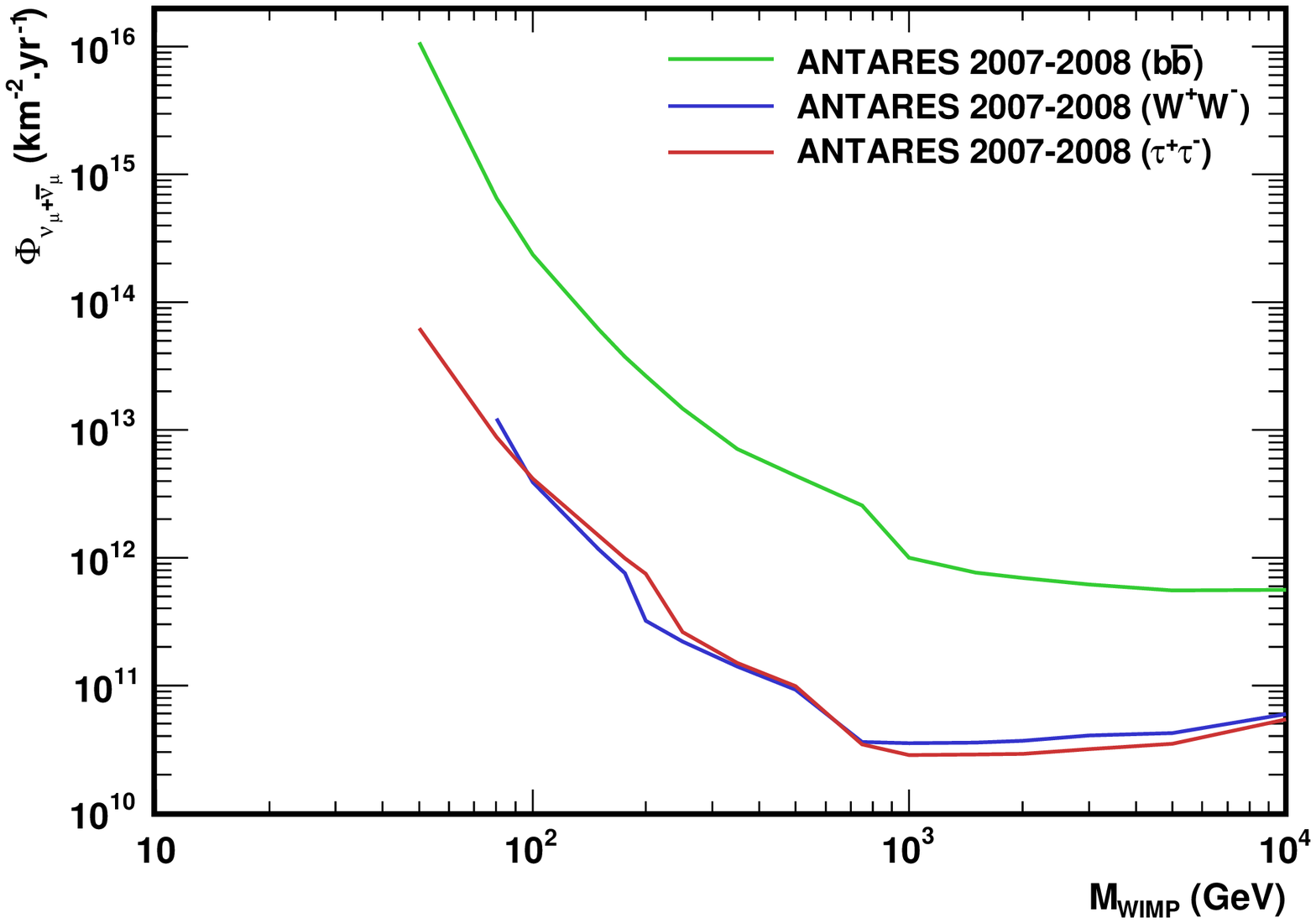}
\includegraphics[width=0.48\linewidth]{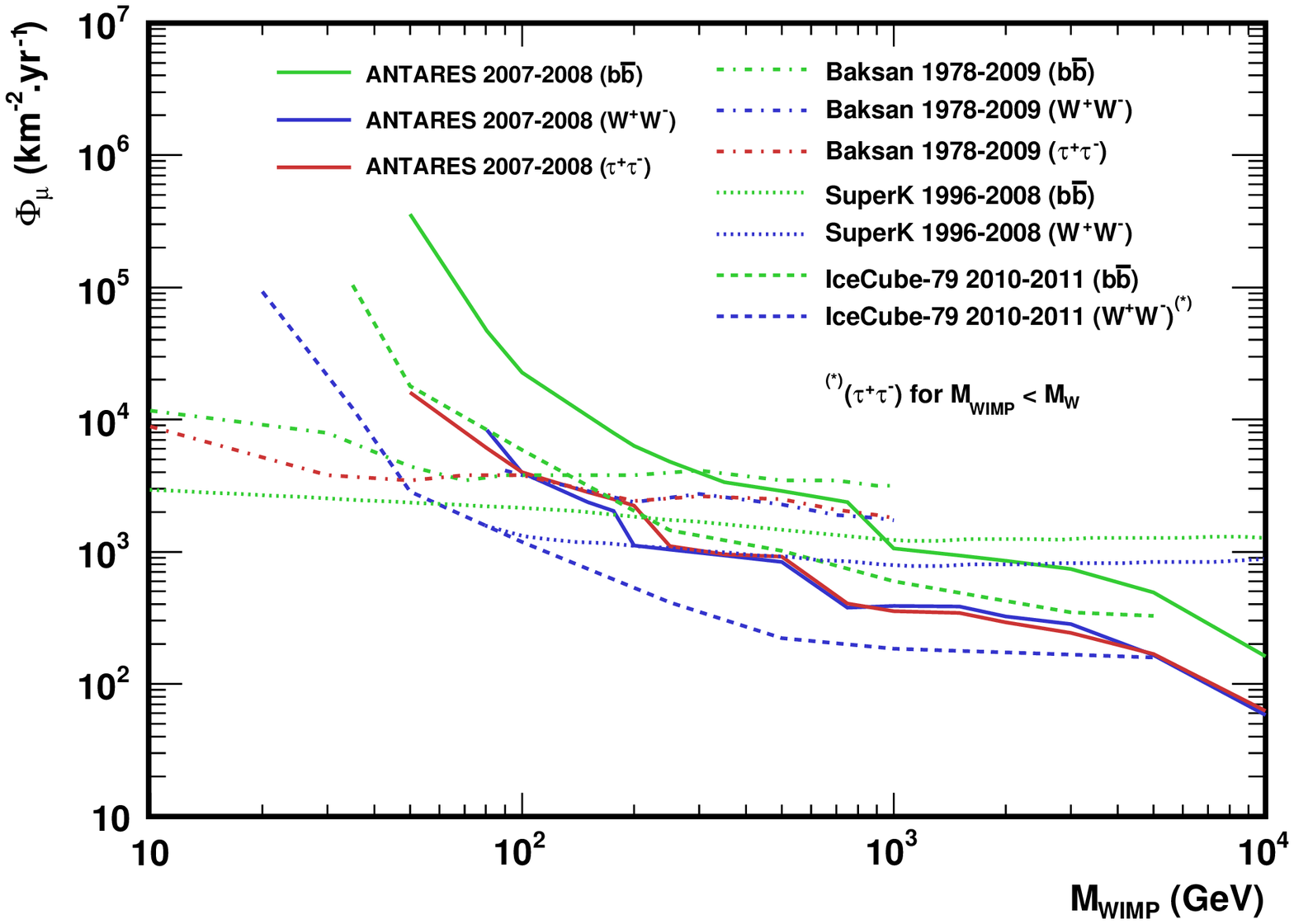}
\caption{Left: $90$\% CL upper limits on the neutrino plus anti-neutrino flux 
as a function of the WIMP mass in the range $M_{\rm WIMP}\in$[$50$ GeV;$10$ TeV] for 
the three self-annihilation channels $b\bar{b}$ (green), $W^{+}W^{-}$ (blue), $\tau^{+}\tau^{-}$ (red). 
Right: $90$\% CL upper limit on the muon flux as a function of the WIMP mass 
in the range $M_{\rm WIMP}\in$[$50$ GeV;$10$ TeV] for the three self-annihilation channels 
$b\bar{b}$ (green), $W^{+}W^{-}$ (blue) and $\tau^{+}\tau^{-}$ (red). The results 
from Baksan $1978-2009$~\cite{baksan} (dash-dotted lines), Super-Kamiokande $1996-2008$~\cite{superk} 
(dotted lines) and IceCube-$79$ $2010-2011$~\cite{icecube} (dashed
lines) are also shown. (Preliminary).}
\label{flux}
\end{center}
\end{figure*}

Once the selection cuts have been optimized, the signal region is
looked at. As shown in Figure~\ref{data}, no excess over the
background has been found, so upper limits in the neutrino flux can be
set, as shown in Figure~\ref{flux} (left). The DarkSUSY
package~\cite{darksusy} allows to calculate the conversion factor needed
to translate the neutrino flux limit into a muon flux limit, shown in
Figure~\ref{flux} (right).

\begin{figure*}[c]
\begin{center}
\includegraphics[width=0.48\linewidth]{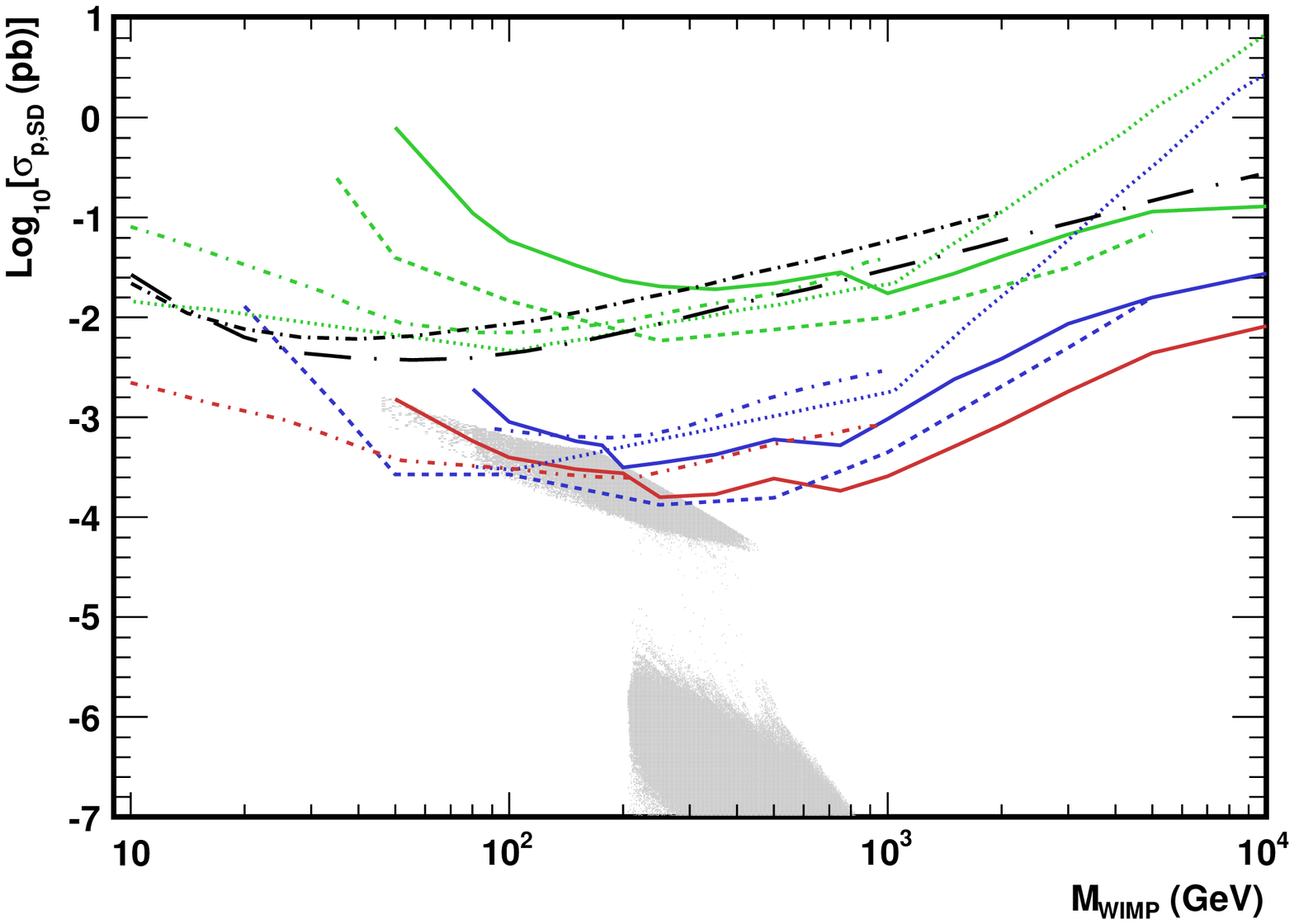}
\includegraphics[width=0.48\linewidth]{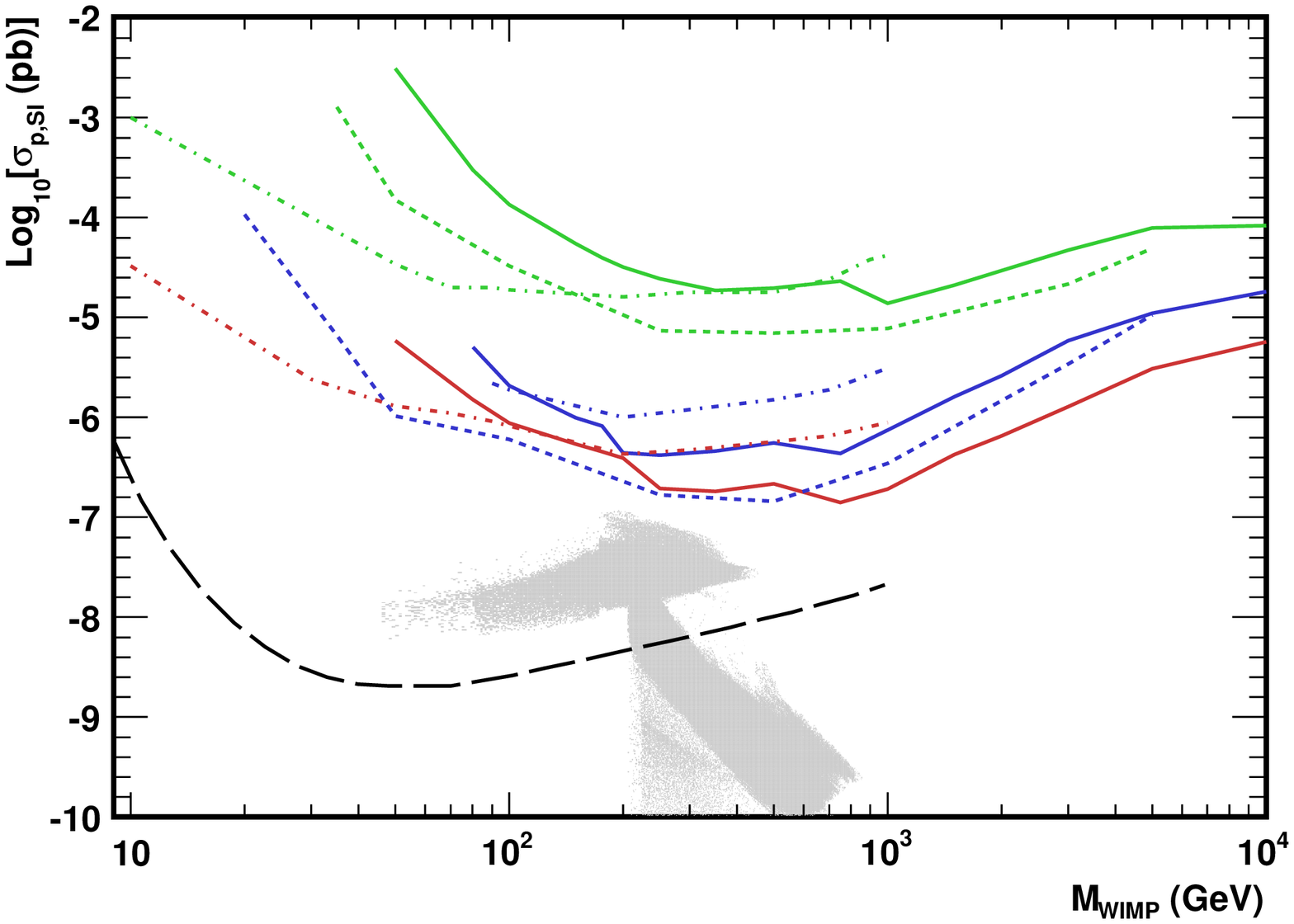} \\
\includegraphics[width=0.48\linewidth]{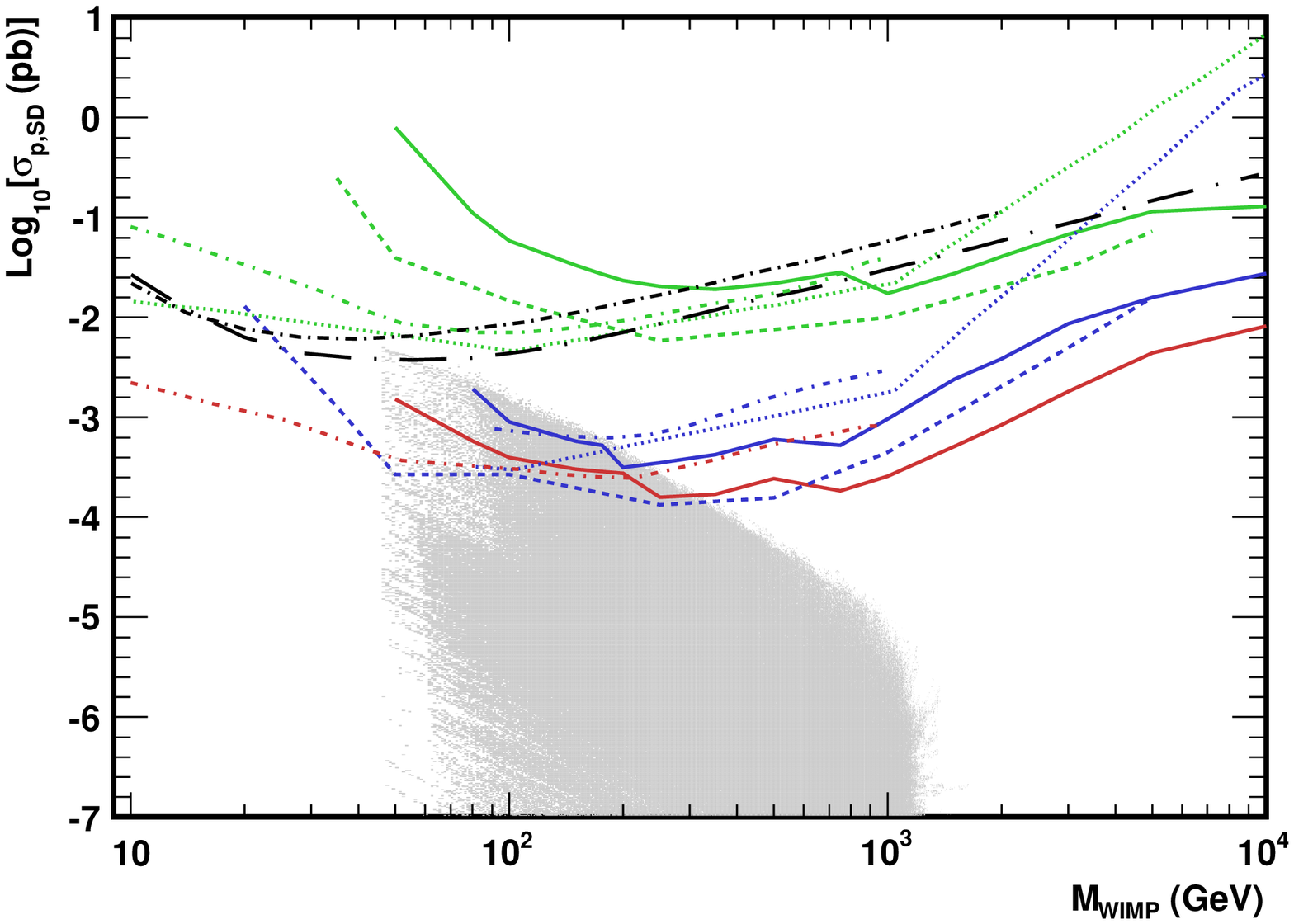}
\includegraphics[width=0.48\linewidth]{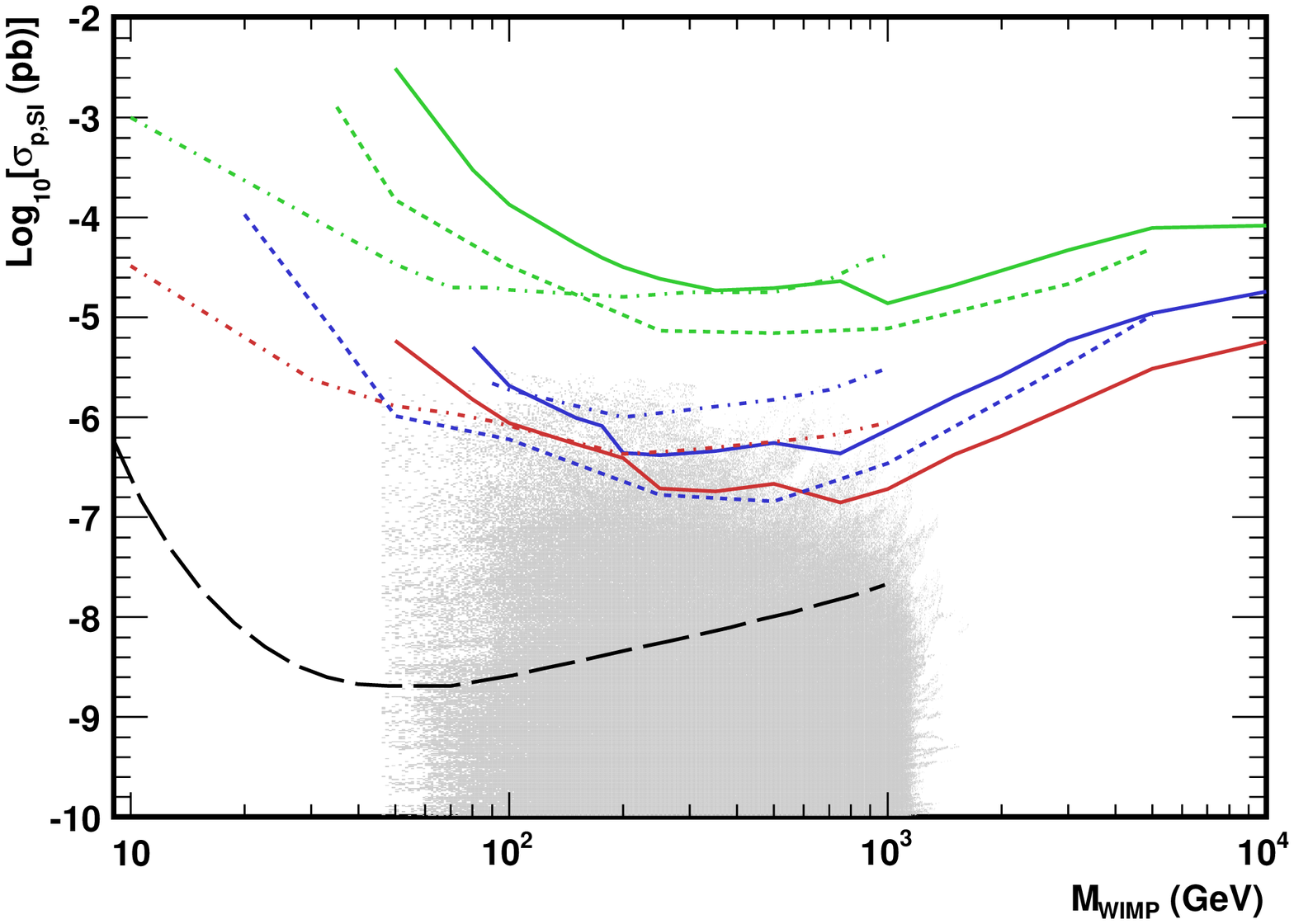}
\caption{$90$\% CL upper limits on the SD and SI WIMP-proton
  cross-sections (plots on the left and right, respectively) 
as a function of the WIMP mass, for the three self-annihilation channels: 
$b\bar{b}$ (green), $W^{+}W^{-}$ (blue) and $\tau^{+}\tau^{-}$ (red), for ANTARES 2007-2008 (solid line) compared 
to the results of other indirect search experiments: Baksan $1978-2009$~\cite{baksan} (dash-dotted lines), 
Super-Kamiokande $1996-2008$~\cite{superk} (dotted lines) and IceCube-$79$ $2010-2011$~\cite{icecube} (dashed lines) 
and the result of the most stringent direct search experiments (black): SIMPLE $2004-2011$~\cite{simple} 
(short dot-dashed line in upper plot), COUPP $2010-2011$~\cite{coupp} (long dot-dashed line in upper plot) and XENON100 
$2011-2012$~\cite{xenon} (dashed line in lower plot). The results of a
grid scan of the CMSSM (upper plots) and CMSSM-7 (lower plots) are included 
(grey shaded area) for the sake of comparison. (Preliminary)}
\label{final}
\end{center}
\end{figure*}

Assuming equilibrium between the neutralino capture rate and the
self-annihilation in the Sun, limits on the spin dependent (SD) and
spin independent (SI) cross section of the WIMP-proton scattering can
be set for the case in which one of them is dominant. These limits are shown
in Figure~\ref{final} compared to different experimental limits and
with the parameter space derived from the CMSSM and MSSM-7 models,
where the lastest constraints from accelerator
experiments have been included (mass of
the Higgs $M_{h}$=125$\pm$2 GeV~\cite{higgs}. A constraint in the neutralino relic
density as 0$<$ $\Omega_{CDM}$ h$^2$ $<$ 0.1232 has also been included.

\section{Conclusions}
\label{conclusions}

The ANTARES data corresponding to 2007-2008 have been used to search
for an excess of high energy neutrinos in the Sun's direction, which
could indicate annihiliation of dark matter particles like neutralinos. The
analysis is a binned search that has shown no excess with respect to
the expectations from background. Upper limits both in the neutrino and
muon flux have been set. Assuming that equilibrium between capture and
annihilation has been reached in the Sun, these limits can be
translated into limits in the spin dependent and spin independent
cross section of the WIMP-proton scattering. The results for spin
dependent cross section are particularly competitive with respect to
direct search experiments. A comparison with the parameter space
allowed by the CMSSM and MSSM-7 models has been shown.

\section*{Acknowledgements}
The authors acknowledge the financial support of the Spanish Ministerio de Ciencia e Innovaci\'on (MICINN), grants FPA2009-13983-C02-01, FPA2012-37528-C02-01, ACI2009-1020, Consolider MultiDark CSD2009-00064 and of the Generalitat Valenciana, Prometeo/2009/026.

\end{document}